\documentstyle[aps,prd,graphicx,floats]{revtex} 

\draft
\def\o{$\cal O$ }
\def\dm{density matrix}
\def\T{temperature }
\def\qms{quantum mechanics }
\def\qml{quantum mechanical }
\def\qm{quantum mechanics}
\def\de{decoherence }
\def\me{ L.Stodolsky}

\def\wf{wavefunction }
\begin{document}
\title{%
\hbox to\hsize{\normalsize\rm 
\hfil Preprint MPI-PhT/2003-11}
\vskip 36pt  COHERENCE AND THE CLOCK }

\author{L.~Stodolsky}
\address{Max-Planck-Institut f\"ur Physik 
(Werner-Heisenberg-Institut),
F\"ohringer Ring 6, 80805 M\"unchen, Germany}

\maketitle
\bigskip
\centerline{ Presented at {\it Time And Matter},  Venice, August
2002}

\begin{abstract}
 We discuss the notion of quantum mechanical coherence in its
connection with time evolution and stationarity. The transition
from coherence to decoherence is examined in terms of an equation
for the time dependence of the density matrix. It is explained how
the decoherence rate parameter  arising in this description is
related to the ``unitarity defect'' in interactions with the
environment as well as the  growth in entropy of the system.  
Applications to the ``Zeno-Watched Pot Effect'' and gravitational
interactions are given. Finally, some recent  results on
applications to macroscopic coherence with the rf SQUID, where the
transition from quantum to classical behavior could be studied
experimentally, are shown.      
 
\end{abstract}
\vskip2.0pc

There's no doubt that the concept of time is  absolutely
fundamental, be it for philosophers, physicists or the  man and
woman on the street. 
Kant even snubbed those haughty mathematicians and put time before
number, making the intuition of time the
origin of arithmetic~\cite{hersh}.  Peculiar to physicists
however, and
probably not agreed upon or well  understood by these other
demographic groups,
is that with the 20th Century lessons of relativity  and \qms
behind
them,  physicists in their overwhelming majority will agree that
time should be viewed as a physical process. There seems
to be no avoiding--even in principle-- the need to understand
how the clock works as a physical object. 

 We are familiar with the slowing down  of time in special
relativity, perhaps most easily understood as a transformation of
fields due to velocity. The various fields change in such a way,
according to relativity,  that any  physical process--including
some
``clock''-- goes slower. Similarly, gravitational fields can affect
the clock, as in the red shift.

While these are interesting enough, there are further
and perhaps less familiar  aspects
of  time as a physical process resulting from quantum mechanics.
This will be the subject of our
discussion.

\section{Time Evolution as a Coherence Phenomenon}
 For a clock to work, according to \qm , it is necessary to have A)
 different 
 energy states  present, and B)  coherence between these states.
This might be easier to understand if we
first look at the more familiar situation for space instead
of time. In order to
have a highly accurate definition of a point in space we need,
according to \qm , a well localized wavepacket. This contains a
large spread of momentum states. Indeed if we had only {\it one}
momentum
component the \wf would be completely spread out everywhere and in
no definite position at all.
 So  localization in position requires a spread of momentum
states.

 But this is not enough. In addition these momentum states must be
{\it coherent}. If one momentum state alone is spread out
everywhere, it
is only the phases between the different momentum states that
allows them to cancel everywhere but in the vicinity of one point.
And so if these states are added {\it incoherently} the system 
is still  not localized.

 And so is it also with energy eigenstates and time. A single
energy eigenstate by itself does not tic, it must ``beat''
or interfere with something else. Typically, with atoms or
molecules, we have a second energy eigenstate nearby and we observe
the oscillating electromagnetic field resulting from the beating of
the two levels.  Or we can suppose a $K^o$-meson
clock~\cite{kclock} or
neutrino clock using the oscillations in flavor (strangeness,
electron-muon number) resulting from the 
interference of different mass eigenstates. If the two atomic or
molecular levels are not coherent, the emitted photons will be
``chaotic'' and the electromagnetic field has no well-defined phase
suitable for a clock; if the two kaon or neutrino states~\cite{nu}
are incoherent there will be no oscillations.

 We thus see that in addition to the slowing down of time which is
possible by relativistic effects, there is a further aspect to time
as physical process: coherence and incoherence.

\section{density matrix}
 To deal with such questions of coherence and
incoherence in \qms it is  not enough  to think about the
wavefunction of the system in question. We need a quantity which
can
describe something like an average over different wavefunctions.
 
 The tool for this in \qms  is the density matrix. The \dm ~is easy
to understand if we suppose having--in our minds--the complete, big
\wf of the whole world, or at least that part of it that might
interact with what we want to study. What we want to
study will in general only be a subsystem of what is described  by
the big \wf and  involves only  a few of the total number of 
variables. Let us take the simple case where this subsystem is a
two-state system, with basis states $\phi_1, \phi_2$. (One state
would be {\it too} simple.)
Then the big \wf can be written
\begin{equation}\label{psi}
\Psi= |x_1,x_2.....x_N;1>\phi_1 + |x_1,x_2.....x_N;2>\phi_2\; ,
\end{equation} 
where the $x$'s are all the variables outside the subsystem, with
$N$ a big number like $10^{23}$, and the kets $|....>$ are the
wavefunctions for these variables. The ``1'' or ``2'' indicate that
these in general depend on the state of the subsystem. Now if we
only look at the subsystem and not at the $x$'s,
that is are only concerned with operators that act on 
$\phi_1, \phi_2$, then all we need to know is the ``reduced''
density matrix of the two-state system. This is a 2 x 2 matrix 
which can be expressed~\cite{us}
 as follows
\begin{equation}\label{rhomat}
\rho=\pmatrix{<x_1,x_2.....x_N;1|x_1,x_2.....x_N;1>&<x_1,x_2.....
x_N 
;1|x_1,x_2.....x_N;2>\cr<x_1,x_2.....x_N;2|x_1,x_2.....x_N;1>&<x_
1,x_2.....x_N;2|x_1,x_2.....x_N;2> }
\end{equation}

where the interpretation of this matrix can be given as:

\begin{equation}\label{pmat}
\pmatrix{ \rm{how~ much~ of~ 1}& \rm{coherence~ between~ 1~ and~
2}\cr \rm{(coherence~ between~ 1 ~and~ 2)^*}& \rm{how~ much~ of~
2}}\; ,
\end{equation}
where of course the ``amounts'' are probabilities and are
normalized to one: ( $\rm{how~ much~
of~
1}+\rm{how~ much~ of~ 2}= 1$). The justification for calling the
off-diagonal elements the degree of coherence between the two
states may be seen by considering two extreme cases: let 
$|x_1,x_2.....x_N;1>$ and
$|x_1,x_2.....x_N;2>$ be either the same or orthogonal. In the
first case $\phi$ can be factored out in Eq~[\ref{psi}] and $\Psi$
is just a product
wavefunction; $\phi_1$and $ \phi_2$ are completely coherent. In the
second case
no operator working just on the $\phi_1, \phi_2$ subspace can cause
the two
states to interfere, they are completely incoherent. If we think of
the $x$'s as describing an environment in which our subsystem  is
immersed, we can say that to the degree the subsystem
is correlated with the environment, the subsystem is internally
incoherent. 

Finally,
this environment could be a ``measuring apparatus''. This is
something that establishes a correlation--usually a strong one--
between the state of the subsystem and the 
$|x...>$, which now represent the ``apparatus''. In the limit of
very strong correlation  we can know what
$\phi$ is just by
looking at the $|x...>$, which is of course how a measurement
instrument should function. On the other hand, we thus see how the
measurement
``collapses the \wf '' of the subsystem: $\phi_1$
and $\phi_2$ become incoherent~\cite{us} when the $|x...>$ states
are very different.

\section{thermal state, stationarity}
 An easy and limiting case of incoherence to understand  is the
thermal
state.
The density matrix operator $\rho$ for a thermal state is
proportional to 
\begin{equation}\label{boltz}
\rho \sim e^{-H/kT}\;,
\end{equation}
where $H$ is the hamiltonian operator, $T$ the temperature and $k$
the
boltzmann constant.
Looking at this expression you might be inclined to  see only the
usual boltzmann factor of statistical physics, which it is.
However, more
important for our discussion is the fact that it is an {\it
operator}. Furthermore this operator is a function of the
Hamiltonian operator only, $T$ and $k$ being just numerical
parameters. This means that there is a definite basis in
which $\rho$ is diagonal, namely  the basis of energy eigenstates
$|E>$: 

\begin{equation}\label{del}
\rho_{E,E'} \sim C_E \delta_{E,E'} \;.
\end{equation}

This diagonal property has the physical interpretation that
different energy eigenstates are {\it incoherent}. That is,
consider the
expectation value of some operator \o :
\begin{equation}\label{tr}
\overline{\cal O}=Tr\bigl[\rho {\cal O}\bigr]\sim C_E
\delta_{E,E'}{\cal
O}_{E,E'}\sim C_E {\cal O}_{E,E} \;,
\end{equation}
In  words, the contributions to ${\overline{\cal O}}$ for each
energy state  are to
be added separately and incoherently.

 One obvious consequence of this incoherence between energy states
is that nothing can be time dependent: as we well know the thermal
state is {\it stationary}. That is, if we consider the time
dependence of
any operator ${\cal O}(t)= e^{iHt}{\cal O}e^{-iHt}$ in the thermal
state we get
\begin{equation}\label{tra}
\overline{\cal O}(t)=Tr\bigl[\rho~ e^{iHt}{\cal O}e^{-iHt}\bigr]=
Tr\bigl[\rho{\cal O}\bigr]= constant \;,
\end{equation}
since we can use the fact that $ e^{iHt}$ commutes with $\rho$ and
the permutation property of the trace to bring the $ e^{iHt}$
factors together and cancel them.
 
 Is the thermal state the only state like this? Obviously not, for
Eq~[\ref{tra}] we see that the important point is Eq~[\ref{del}],
the diagonality, and not 
 the precise value of the $C_E$. The above arguments go through for
{\it any} $\rho= f(H)$ and not just Eq~\ref{boltz}. For this reason
we
wrote a general coefficient $C_E$  and not the boltzmann factor. So
if it's not  the thermal state, what is the characterizing feature
of the \dm~[\ref{del}]? The property  which will be true of any
state describe by  Eq[\ref{del}] is that it is {\it stationary}.
Since there is no
coherence between energy states no operator can show a time
dependence.

  Conversely, if we know on physical grounds that some system is
stationary, we know that its energy states are incoherent. An
important case of this is the radiation coming from a thermal
source. The radiation itself, say neutrinos from the sun, need not
follow a boltzmann factor. But since the sun, (to a very high
degree of approximation) is stationary, we know that different
energy states are incoherent. Understanding this helped to clear up
a long-standing confusion on the treatment  of
neutrino oscillations~\cite{unn}.

\section{loss of coherence} 
Limiting cases like  perfect coherence  for 
a pure state, or incoherence of  energy states
for
stationary systems, are easy to understand. However, we
are often
confronted with the more difficult
situation of partial and time-dependent coherence. Typically we
imagine the problem of
a subsystem 
 in a pure, perfectly coherent, state at $t=0$; which as time
goes on  becomes more incoherent as it interacts with the
external world. Although this is in principle a perfectly well
defined
problem: just go ahead, find the total \wf at any time and average
out the unobserved variables by calculating Eq~[\ref{rhomat}]
---there is no
simple way to do this perfectly rigorously for a many-variable
system.   We must resort to some kinds of plausible
approximations or guesses.

 A simple and reasonable picture, from which however we can learn
a
lot, amounts to assuming that the outer world or environment has
steady properties with time and there is a loss of coherence at a 
 constant  rate for our subsystem. Formally this is like the
``golden rule'' calculation used in decay problems or scattering
theory.  For the two state system as described by Eq~[\ref{rhomat}]
we can give a quite complete description of the phenomenology of
the problem on this basis and draw a number of interesting
conclusions.

\section{parameterization of $\rho$}
To set up the framework for this approach, we first parameterize
the 2x2 matrix for $\rho$ in terms of the familiar pauli matrices
$\bf
\sigma$:
\begin{equation}\label{rhomata}
\rho= {1\over 2}(I +{\bf P}\cdot {\bf \sigma}) \; ,
\end{equation}
which is appropriately normalized to $Tr \rho =1$.
The two-state subsystem is governed by an internal hamiltonian
which completely describes its behavior in the absence of
interaction with the environment. It is also a 2 x 2 matrix which
we can parameterize in the same way:

\begin{equation}\label{hmat}
H= {\bf V}\cdot {\bf \sigma} \; ,
\end{equation}
where we have left out a possible constant term which simple
represents an overall energy shift and has no interesting effects.

These are just mathematical definitions, but because of the
familiar
identification of the pauli matrices with spin, they allow a
simple intuitive picture. The 3-vector $\bf P$ is like a
``polarization''. According to  Eq~[\ref{pmat}] its vertical or
``z'' component represents the relative amounts of the two states
``1'' or  ``2''. The ``transverse'' or x-y
components, according to Eq~[\ref{pmat}]  represent the degree of
coherence within the subsystem. In generalizing to more than two-
state systems, this remains true; the diagonal elements represent
the amount of the ``quality" in question and the off-diagonal
elements
the coherence between them. Note that this is necessarily
a basis-dependent picture: we must specify which ``quality'' we are
talking about.  

\section{Isolated system}
 Naturally if the system is isolated and has no interactions with
the outside world the situation is very simple.  The evolution is
governed by the above hamiltonian and the equation of motion for
$\rho$ given by the basic formula

\begin{equation}\label{rhodot}
i\dot \rho=[\rho,H]
\end{equation}
Using the algebra of the $\sigma$ this translates into

\begin{equation}\label{pdot}
\dot {\bf P}={\bf P}\times {\bf V}\; .
\end{equation}
Pictorially, the ``polarization'' {\bf P} precesses around a  
``pseudo-magnetic field'' {\bf V}. Note that {\bf V}
can be time dependent if we wish. So far we haven't done very much,
this is just equivalent to the evolution of a spin in a magnetic
field. Observe that with this equation the length of {\bf P} cannot
change: 
\begin{equation}
\frac{d P^2}{dt}=2  {\bf P}\cdot({\bf P}\times {\bf V})=0\;. 
\end{equation}
That is,  {\bf P} rotates without shrinking. This represents the
fact that a pure state  remains a pure state for an isolated
system.  

\section{environment} 
The non-trivial part comes in when we try to handle the interaction
with the outside world. As mentioned, if we make the assumption of
an outside world with constant properties, interacting repeatedly
with our subsystem such that  there is a steady loss of
coherence, that is a constant rate of decrease of the off-diagonal
elements, we can obtain a plausible generalization~\cite{siegel} of
Eq~[\ref{pdot}]. This has the form

\begin{equation}\label{pdota}
\dot {\bf P}={\bf P}\times {\bf V}- D{\bf P}_{T}\; .
\end{equation}
This involves one new parameter, $D$. This is the damping or
decoherence rate and describes the rate of loss of coherence
between the two basis states. $ {\bf P}_{T}$ means the
``transverse'' or $(x,y)$ components of {\bf P}.  If we mentally
turn off    the first term in the equation, the $D$ term leads to
an exponential decay of the overlap or coherence terms in $\rho$, 
as in the ``golden rule'', and  this rate is independent of the
time
when we start. If our subsystem is immersed in a large essentially
infinite system it is difficult to imagine any other behavior.

 We have made one inessential simplification in writing
Eq~[\ref{pdota}]. In order to concentrate on the interesting
interaction between the decoherence and the internal evolution we
have left out a possible term $D'P_z$ in which there would be a
direct relaxation between the two basis states. For example in the
problem of two states of a chiral molecule separated by a barrier
~\cite{us} we are assuming that the temperature is low enough that
there is no
direct jumping over the barrier. This is the simple classical
relaxation process which can easily be dealt with but is not
essentially relevant to the quantum mechanical aspects on which we
wish to focus.
Similarly, it will be seen that at long times {\bf P} in
Eq~[\ref{pdota}] tends to zero. This may not be exactly correct if
the internal hamiltonian is such that there is a constant energy
difference between the two basis states, large compared to an
ambient temperature. Then $P_z$ should end up being given by the
boltzmann factor and not be zero. So we implicitly make the
assumption that any constant component $V_z$ is small compared to
the temperature.
 
 The most important aspect of the presence of the damping term  is
that now the length of $\bf P$ can shrink:
\begin{equation}\label{shr}
\frac{d P^2}{dt}=2{ \dot P}\cdot{ P}=2{\bf \dot P}\cdot{\bf P}=-2
D {\bf P}\cdot{\bf
P}_T\neq 0 \; ,
\end{equation}
 in
general, and so the density matrix can change its ``purity''.
 That is, even without explicitly ``quality changing'' relaxations
the length of P will inevitably decrease, if there is
a ${\bf P}_T$ component. Such a component however, will be
produced, in general, by the ``rotation'' induced by $\bf V$.

Although we shall explain below how to arrive at a microscopic
understanding of 
$D$, even the simple phenomenological  Eq~[\ref{pdota}] allows us
to arrive at some interesting conclusions. One of these is the
``Zeno Effect''.

\section{Zeno-- Watched Pot effect}
 The Zeno or Watched Pot Effect, perhaps    first proposed by Alan
Turing ~\cite{us}, goes something like this. According to the
``Copenhagen School'''s treatment of the ``Measurement Problem''
(all these things are in ``...'' because I'm not sure they exist ),
 a measurement leaves the state of the system in one  particular
eigenstate of the measurement. For example after a position
measurement, the object is in some definite position--not in a
superposition of positions. So we or Turing might say,  let's keep
rapidly measuring the
object's position. It will repeatedly be fixed to the spot and
not be able to move at all! 

 Sometimes this ``paradox '' is used to say there's something wrong
with \qms or something of the sort. But actually it represents 
perfectly reasonable \qml behavior and in fact is just a solution
to Eq~[\ref{pdota}], for the case of large $D$.

 To see this, consider 
the behavior of Eq~[\ref{pdota}] in the case of a constant $\bf V$,
say along the x-axis. If we start $\bf P$ along the z-axis, it will
tend to rotate in a circle in z-y plane: $P_z$--the probabilities
or ``amounts'' we have defined as represented by  $P_z$
will oscillate.
Now if we turn on $D$, the oscillations will be damped and
gradually
die out. As we make $D$ larger and larger the oscillations will 
become overdamped and disappear completely. Finally in the limit of
very strong damping one finds~\cite{us}, ~\cite{nu}

\begin{equation} \label{zeno}
P_z\sim e^{- { V^2\over D} t} \; .
\end{equation}

In the limit of large $D$, $P_z$ is practically frozen and
will hardly budge from the spot! According to this formula the time
scale, initially set by the oscillation time $1/V$, gets stretched
by a factor  $V/D$, which can be enormous when it turns out $D$
is something like the interaction rate in a gas (see formula for
$D$ below) and $1/V$ is the tunneling time between two states of a
chiral molecule.
 In this way we gave a second  answer to Hund's paradox on the
stability
of the optical isomers~\cite{us} (the first was parity
violation~\cite{beats}). On a practical level this means that in
relaxation processes at low \T where the classical barrier-hopping
type of mechanisms are frozen out and quantum tunneling takes over,
we have ``anti-intuitive'' relaxation, where a bigger $D$ means
{\it
slower} relaxation~\cite{us}.

 You might say the repeated ``measurements'' by the environment 
have fixed the state a la Zeno- Watched Pot but in any
case it's just a result of a
simple solution to  Eq~[\ref{pdota}] and has nothing mysterious 
about it.

 At this point we should warn of a misunderstanding which sometimes
arises. The arguments,  or similar arguments leading to
Eq~[\ref{zeno}] or its equivalent, depend very much on our dealing
with a system with two, or in any case a finite number of, levels.
They do {\it not} apply to a continuous density of levels, as for
the decay of a quasi-stable state into the continuum. This
misunderstanding has occasionally led some people to erroneously
suggest
that the decay of a particle or atomic or nuclear level could be
inhibited
by observing if it has decayed or not. This is obviously silly.
Mathematically the difference between the two cases is that with
two or finite number of levels  $P_z$ has a ``flat-top''
near $t=0$ in the absence of damping. That is,  $P_z\sim
1-(Vt)^2$ for small times;
while for a true decay problem we have $e^{-\Gamma t}\sim 1-\Gamma
t$, a linear
behavior.
This leads to qualitatively different behavior with respect to
``stopping and restarting'' the system as we do when we turn on
the damping to get the Zeno-Watched Pot behavior.

Another nice way of undertanding this (suggested to me by Michael
Berry) is to consider the decay process as the tunneling through a
barrier, as in the Gamow picture of alpha decay. Now, when there is
tunneling through a barrier connecting two {\it discrete} states,
as in the chiral molecule problem~\cite{us}, the origin of the
Zeno-Watched Pot effect may be understood as follows. Tunneling
between two discrete states is a very delicate process and depends
very much on the quasi-degeneracy of the energy  between the two
states. With $E_{split}$ the energy splitting , the tunneling
probability goes as $\sim(\omega_{tunnel}/E_{split})^2$, with
$\omega_{tunnel}$ 
the tunneling energy. Since $\omega_{tunnel}$ is typically very
small, so must  $E_{split}$ be small to have a non-neglible
tunneling probability. Now if some external influence is causing
the energy levels to move around constantly, even by small amounts,
say by some shifting  of the two potential wells in the two-state
tunneling problem, the degeneracy is 
constantly being lifted and the tunneling is slowed down. Depending
on the ratio of $D/\omega_{tunnel}$ it can be practically stopped.
This leads to
the Zeno-Watched Pot effect. 
But  the situation is entirely different if we have a quasi-bound
state decaying into a {\it continuum} of states, as in the alpha
decay situation. Even if the energy of the quasi-bound state is
fluctuating, it will always find some continuum state with which it
is degenerate. Hence there is no Zeno-Watched Pot effect.

\section{Formula for D or the ``Unitarity Deficit''}

 We now come to the microscopic understanding of $D$, the damping
or decoherence rate. Using our general ideas, we can derive a nice
formula for this, which furnishes considerable insight into the
``decoherence'' or ``measurement process".

From its definition in Eq~[\ref{pdota}] and looking at
Eq~[\ref{pmat}] we see that $D$ is a rate parameter, one which
gives the rate at which the overlap between $|x_1,x_2.....x_N;2>$
and $|x_1,x_2.....x_N;1>$ is decreasing with time.
We take our subsystem to be interacting with the big environment in
a constant, steady manner--this is implicit in our assumption of a
constant $D$ with no further time dependence. We can model this
situation by looking at the subsystem as  being bombarded by a 
constant flux from the environment. This may be thought of
as particles, or excitations like phonons or
quasiparticles. Our basic idea is that if the two  states 1 and 2
of the
subsystem scatter this incoming flux in a way which is different
for 1 and 2, then the overlap
$<x_1,x_2.....x_N;1|x_1,x_2.....x_N;2>$ decreases.

 We recall the concept of the 
$S$ matrix of scattering theory. This is an operator which turns
the incoming wavefunction 
into the outgoing wavefunction:

\begin{equation}\label{s}
\psi_{out}= S\psi_{in}\;.
\end{equation}

 Now the important point here is that states 1 and 2 may scatter
the incoming \wf differently, so we have two $S$ matrices, $S_1$
and $S_2$~\cite{soph}. If these are different the incoming particle
or
excitation, which was {\it un}correlated with the state of the
subsystem
before the scattering, will be correlated with it afterwards.
  So every scattering decreases the overlap and if we work out what
this means for $D$ we get the following formula: 
\begin{equation}\label{d}
D=(flux)~ {\mathrm Im}~ i ~<i\vert( 1-S^\dagger_1S_2  )\vert i>\;
,
\end{equation}

 The expression is proportional to the incoming flux since it is a
rate parameter and $\vert i>$ 
refers to the incoming state, with a possible average implied if we
have many kinds of incoming states.
 ``Im '' means imaginary part. There is also a significance to the
real part, which is an energy shift of the subsystem induced by the
environment~\cite{us}. 

Since usually we have $S^{\dagger}S=1$ for the $S$ matrix,
the formula says that the decoherence rate is related to the
``unitarity deficit'' produced by the fact that the different
components of the subsystem don't interact  the same way with
the outside world.

   Eq~[\ref{d}] has two 
illuminating limits:

\begin{equation} \label{no} 
S_1=S_2~~~~~~~~~~~~~~~~~D=0\; ,~~no~ decoherence
\end{equation}
If both states of the subsystem interact equally with the outer
world, $D=0$, there is no decoherence. This is in accord with our
feelings about ``measurement''. If the ``outside'' does not respond
to the differences in the subsystem there is no ``measurement'' and
no ``reduction'' of the state. However, we do not need to use this
language, the equation stands by itself.  
Note an important point: interaction with the environment is
possible without decoherence.

The other interesting limit occurs if one state, say 1, doesn't
interact so $S_1=1$ then
\begin{equation}\label{opt}
S_1=1~~~~~~~~~~~~~~D=1/2 \times (scattering~ rate~on~state~2)
\end{equation}
This result follows from an application of the optical theorem of
scattering theory $(cross~ section) \sim Im ~S$ and
$Rate=(flux)\times (cross ~section)$. This corresponds to the naive
expectation that the decoherence rate is the scattering
rate. This is just one particular limit however, and there can be
subtleties 
such as effects of phases on $D$~\cite{vbl}.

\section{the unitarity deficit and entropy production}
There ia a classical conundrum which says, in apparent
contradiction to common sense, that the entropy  of a isolated
system cannot increase. This is particularly simple to show in \qm 
~where the definition of entropy is $ -Tr[\rho \,ln\rho ]$. Since
for an isolated system the evolution of $\rho$ is governed by a
Hamiltonian, we get for the time dependence of the entropy, just as
in
Eq.~(\ref{tra}), $Tr\bigl[ e^{iHt}\rho\, ln \rho e^{-iHt}\bigr]=
Tr\bigl[\rho\, ln \rho\bigr]= constant$. This argument says that as
long as the time dependence of $\rho$ is governed by a Hamiltonian,
the entropy is constant.

 This constancy would also apply to the entropy  $-Tr[\rho\,ln \rho
]$
of our
two-state system where $\rho$ is given by Eq.~(\ref{rhomata})  
if it were isolated --not coupled to the environment. Or as we
learn from the argument, it could even be  coupled to the
environment; but in such a way that the evolution of $\rho$ 
is given by a  Hamiltonian. However we
see from Eq.~(\ref{d}) that the coupling to the environmnet is not
governed by a single Hamiltonian but rather by two Hamiltonians,
giving the two $S$
matrices. If we had one Hamiltonian we would have  $S_1=S_2$,  in
which
case  $D=0$, and there is no decoherence.

 Hence there is a connection between
$D$ and the entropy increase of the two state system. In fact 
diagonalizing Eq.~(\ref{rhomata}) and taking the trace, we find 
for the entropy $-Tr\bigl[\rho\, ln \rho\bigr]=ln{ 2}-{1\over
2}\bigl((1+P)ln(1+P)+(1-P)ln(1-P)\bigr)$. Taking the time
derivative,
we find for the entropy change of the two state system

\begin{equation}\label{ent}
{d (Entropy) \over dt}=- \dot P P{1\over 2P}ln{(1+P)\over(1-
P)}=(D{\bf P\cdot P_T}){1\over 2P}ln{(1+P)\over(1-P)}\approx D{\bf
P\cdot P_T}\; ,
\end{equation}
 where we used
Eq.~(\ref{shr}). The $\approx $ refers to the limit of small $P$.
It seems intuitively clear that the rate of entropy increase
and the decoherence rate should be closely related and
Eq.~(\ref{ent}) expresses this  quantitatively. By appropriate
generalization of the various quantities, this could be extended to
 systems larger than just two states. Note ${\bf
P\cdot P_T=P}_T^2$ is necessarily positive. Furthermore, in thermal
equilibrium where there is no coherence between states of
different energy i.e
${\bf P}_T=0$,  there is no entropy production.
    
\section{Decoherence in mesoscopic devices}
 In recent years our subject has moved from the theoretical-
philosophical to the domain of the almost practical with the
realization  of quantum behavior for large, essentially macroscopic
devices--``mesoscopic systems''. This has been given additional
impetus in view of the possible  use of such devices for the
implementation of the ``quantum computer''.  Furthermore the
subject is  interesting in connection with the idea--to my mind
wrong--that there might be some  limit where large objects don't
obey the rules of quantum mechanics. Decoherence is of course the
main question for the observability of such effects and their
possible practical use.

One of the devices that has been studied in detail is the rf SQUID,
where
by suitable adjustment of parameters it is possible to bring the
system into the configuration  of the familiar  double potential
well separated by a tunneling barrier. The ``x'' coordinate  stands
for the flux in the superconducting ring, and the system obeys--in
the absence of decoherence-- a Schroedinger equation in this
variable. The  states localized in one of the  potential wells 
represent the supercurrent flowing in a given sense around the
ring, with a very large number (microamps) of electrons reversing
direction when we go from one well to the other. Creation of
quantum linear combinations of these states, which can occur by
tunneling, would certainly be impressive evidence for the general
applicabilty of quantum mechanics. Some beautiful experiments
~\cite{beau}~\cite{beau1} using microwave technique have seen
evidence for such combinations in SQUID systems.

 We have suggested~\cite{squid} a particularly simple way to both
see the
quantum linear combination
of the two states and to measure the decoherence time of the
system, hopefully  directly sensitive only to the true quantum
decoherence.  It involves the idea of ``adiabatic inversion''. This
occurs when a slowly varying external field can cause a quantum
system to reverse its state, as when a spin ``follows'' a rotating
magnetic field and goes  from up to down. This phenomenon is also 
often refered to as a ``level crossing''. It  is an intrinsically
quantum mechanical phenomenon
and, --important for us-- is hindered when the decoherence time is
short compared to the time in which the  inversion takes place.

We  propose to produce such an inversion in the  SQUID by sweeping
an external field and then  observing the reversal of the direction
of the flux in the  SQUID. Note that the system need not be
``observed'' until the procedure is over---our method is
``noninvasive''. When the sweep is faster
than the
decoherence time the inversion should take place, and when it is
slower it should be inhibited. We are witnessing the transition
from
quantum (tunneling allowed) to classical (tunneling forbidden)
behavior as the  decoherence is increased. 

 Going  from fast to slow sweeps, the sweep time where the
inversion begins to become unsuccesful thus gives a determination
of the
decoherence time. A possible difficulty here is that the sweep
cannot be too fast, otherwise the procedure becomes non-adiabatic.
However, our estimates indicate that a region of SQUID parameters
and temperature should exist where fast-enough  sweeps are possible
without violating adiabaticity. 

 In order to study these points in more detail we have developed
numerical simulations of such systems, both for the study of the
adiabatic inversion (the logical NOT) as well as for a two-SQUID
system operating as a CNOT quantum logic gate~\cite{cnot}. In Fig~1
we show the results of a simulation~\cite{sim} for the one-SQUID
adiabatic inversion. The  decoherence
time $1/D$ was chosen to be about 39 000 units and simulated    as
a random flux noise. The SQUID parameters were $\beta=1.19,
L=400\,pH, C=0.1pF$, giving a time  unit of $6.3\times 10^{-12}s$
and so  $1/D= 0.25 \mu s$ . This   decoherence
time would correspond to about $T=50\,mK$ in the estimate $D=T/(e^2
R)$, with $R=5M\Omega$~\cite{squid}.   The simulation included the
first 8
quantum levels of the SQUID so that possible effects of non-
adiabaticity are taken into account.

The vertical axis in 
Fig 1 shows the probability for finding the flux in the
 ring reversed  after the sweep. We see that while the
inversion is essentially always successful for sweep times less
than decoherence time, it becomes progressively less so for longer
sweeps. Hence we are seeing  the transition from quantum towards
classical behavior, and a measuremnt of when this takes place
furnishes a determination of the decoherence time.
   The gradual fall-off seen
after the completion of the longest sweep is indicative of another
difficulty, relaxation. Our final states will in general not be
distributed according to thermal equilibrium, and  the final
detection should take place
quickly on the time scale for relaxation.

\section{decoherence and gravity}

Although the above ideas are rather simple and phenomenological,
they can be applied to a wide variety of interesting problems,
where of course the various quantities  like $\bf V$ and $S_1, S_2$
must be adapted to each particular case. These range from
understanding the permanence of optical or ``chiral'' isomers
(Hund's paradox)~\cite{us}, to the study of neutrino
oscillations~\cite{nu}, to the
mesoscopic
devices ~\cite{vbl},~\cite{squid}, and the design of
quantum logic gates~\cite{cnot} just discussed.

 Here, however, I would like to conclude in a more speculative
vein,
coming back to the question of the need for coherence between
energy states to define time. Is the loss of coherence always a
more or less accidental happening, depending on the particulars of
the situation? Or is there something universal about it?

There is of course one kind of interaction which universally
and inevitably
 couples to energy states and which in principle 
``measures'' or distinguishes them: gravity~\cite{grav}. Gravity
couples to
mass/energy and so states of different mass/energy interact
differently. Indeed one can derive the gravitational redshift by
considering the  $K^o$ meson ``clock'' ~\cite{kclock}. 

For this clock,
we have two mass/energy eigenstates and interference effects
between them  tell time.
 By the same token this must imply some degree of decoherence
between the different mass/energy states due to gravitational
interactions. There ought to be an imaginary or dispersive part to
the redshift. Naturally because the coupling, Newton's constant, is
so small  we expect the effects to be negligible under everyday
conditions. It is nevertheless amusing   to see what
the formulas look like. We shall incidentally find that
there is in fact an ``everyday'' application, involving the passage
of oscillating neutrinos near galaxies.

Qualitatively, we expect  in regions of large and rapidly varying
gravitational fields that different energy states become
incoherent. Since gravity couples universally, this cannot be
dismissed as an incidental effect. It will affect all energy
states,
and thus time  is slowed down equally for all processes. If the
effects become
strong, time in some sense becomes ill-defined or stands still.

We try to model the \de due to gravity by calculating $D$ for  the
$K^o$
clock. There  are  two interfering mass eigenstates  and  an
environment
interacting gravitationally with these two states. If we take this
environment to be a flux of particles, we can use our formula
Eq~[\ref{d}]. The calculation of the S matrices is (non
relativistically) the same as  for coulomb scattering, where
the ``charge'' is the mass. The two components of the clock have
different masses $M_1, M_2$ and so $S_1, S_2$  are different and
thus
there is  a contribution to $D$. In an impact parameter
representation we have

\begin{equation}\label{imp}
D= (flux) \int 2\pi b~ db~
 Im~i (1-S^{\dagger}_{M_1}(b) S_{M_2}(b)) 
\end{equation}
with $S(b)=e^{2i\delta(b)}$ and 

\begin{equation}\label{sb}
\delta(b) =  2 \int^{l_{max}}_0 \alpha {dl\over                 
\sqrt{l^2+b^2}}\;,
\end{equation}
where $\alpha=GEM/v$, with G the Newton constant and $E$ the energy
of the incoming particle. As is usual in coulomb problems there is
a divergence at large distances, 
and so we introduce a large
distance cutoff $l_{max}$ which must be interpreted according to
the physical conditions. Taking the two $S$'s and the imaginary
part we finally get

\begin{equation} \label{alphad}
D\approx (flux)l_{max}^2(\Delta \alpha)^2\;,
\end{equation}
with $\Delta \alpha=G(\Delta M) E/v$, which we have taken to be
small and expanded to lowest order. $\Delta M$ is $(M_1-M_2)$.
 We can try to estimate this under thermal conditions, as in the
early universe, where all dimensional quantities are related to the
temperature $T$. We end up with
\begin{equation}\label{difin}
D\approx T^3 (G\Delta M)^2 =T^3 ({\Delta M\over M^2_{planck}
})^2\;.
\end{equation}
As would have been expected, 
T must be at least on the order of the planck scale for this to be
significant. And then the \de rate is  $\Delta M (\Delta M/
M_{planck})$, which  is still small, unless we are considering
$\Delta M$  also on the planck scale. The
result Eq~[\ref{difin}] is of order $1/M^4_{planck}$ since
decoherence,  as explained, is a
unitarity effect and so comes in to second order in  the coupling.
Although conceptually interesting this seems very remote from any
potentially observable effects.

Nevertheless there is, as promised,  a ``present-day" 
application. We now know there are neutrino oscillations.
So possibly if the time scale for the release of the neutrinos in
the Big Bang were short compared to their oscillation time (then),
they would still be oscillating in step today. Instead of the $K^o$
clock we have a neutrino clock. A beautiful clock--if we
could ever read it--a precise timepiece going back to the very
first
minutes of the Big Bang. But there's a difficulty:  in traveling to
us the neutrinos will pass
near various mass and  gravitational field inhomogeneities, like
galaxies, and these will tend to mix up the phases.
Because of  the huge mass of a galaxy, $\Delta \alpha$ will no
longer be small. One finds~\cite{grav}, with typical mass
parameters for  neutrinos and galaxies,  that essentially any
neutrino traveling through or near a galaxy will be decohered.
Unfortunately, there will be no oscillations to read. 
So much for the Big Bang Clock, but at least there's
 a  ``practical'' application of \de theory involving gravity.

Note that these ideas, based on the different interaction of
different mass/energy states, do not contradict the classical
equivalence principle according to which different masses move in
the same way. The interaction is so cleverly constructed that while
different masses follow the same path classically, they get
different quantum phases~\cite{kclock}, and these are essentially
what  concern us here. The interplay between the classical and
quantum is subtle and interesting. The last has certainly not been
said on these questions.

\begin{figure}[p]
\includegraphics[height=.85\hsize,angle=-90]{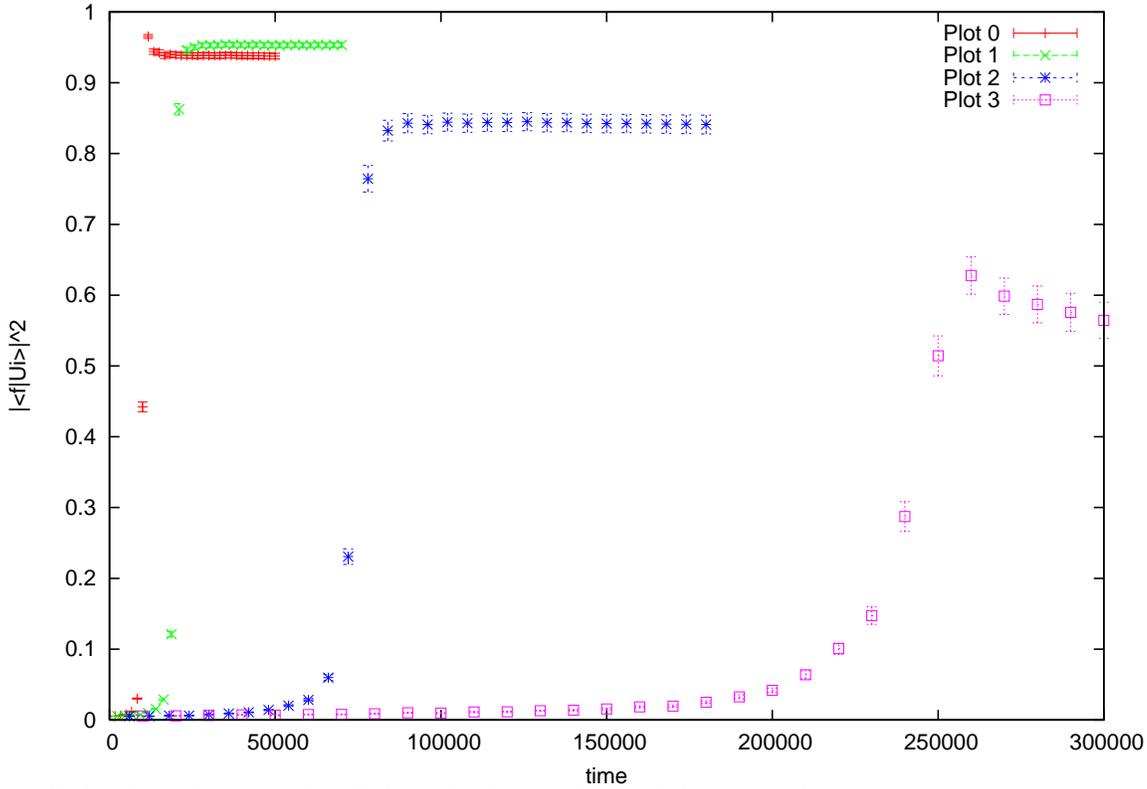}
\caption{ Simulations of  adiabatic inversion in the rf SQUID for
sweep times long and   short compared to the decoherence time. The
vertical axis is the probability for finding the flux in the SQUID
reversed. The simluation 
parameters have been chosen so that decoherence
time  is about $1/D=39\,000$ .  For  sweep times  less than $1/D$
the inversion is essentially completely succesful,
while for longer sweep times it becomes less and less so . An
experiment of this type would show the transition from quantum to
classical behavior and permit a  determination of
the decoherence time.     For the SQUID parameters given in the
text the time interval corresponds to $6.3\times 10^{-12} s$, so
the decoherence time is about $0.25\,\mu \,s$. }
\end{figure}


\begin{references}
\bibitem{hersh}{\it What is Mathematics Really?} R. Hersh, Oxford
(1997).


\bibitem{kclock} Matter and Light Wave Interferometry in
Gravitational 
Fields, \me, Gen. Rel. Grav. {\bf 11}, 391 (1979).


\bibitem{nu}On the Treatment of Neutrino Oscillations in a
   Thermal Environment, \me,  Phys. Rev. D 36(1987)2273 and
 chapter 9 of G.G.~Raffelt,
{\it Stars as Laboratories for Fundamental Physics} 
(Univ. Chicago Press, 1996).
 
\bibitem{us} Two Level Systems in Media and `Turing's Paradox',
     R.A. Harris and \me, Phys. Let. B 116(1982)464.
Quantum Damping and Its Paradoxes,\me,
in {\it Quantum Coherence}, J. S. Anandan ed.
World Scientific, Singapore (1990).

\bibitem{unn} When the Wavepacket is Unnecessary, \me, Phys. Rev.
{\bf D58}  036006, 1998 and www.arxiv.org, hep-ph/9802387.

\bibitem{siegel} See ref 4, and for a more formal derivation, 
focused however 
on neutrino applications:
``Non-Abelian Boltzmann Equation for Mixing and
Decoherence''
 G. Raffelt, G. Sigl and \me, Phys Rev Let.
70(1993)2363.

\bibitem{beats} Quantum Beats in Optical Activity and Weak
Interactions,
    R.A. Harris and \me, Phys Let B78 (1978) 313.

\bibitem{soph} The sophisticated reader will recognized that
although a very simple and intuitive picture arises through the use
of the S matrix with  distinct scattering events, one could reach
a similar result by using a hamiltonian with a continuous time
evolution and depending on a  quantum number with values ``1'' and
``2''.

\bibitem{vbl} Measurement Process In a Variable-Barrier System,\me,
Phys. Lett. {\bf B459} 193, (1999).

\bibitem{beau}J. R. Friedman, V. Patel, W. Chen, S. K. Tolpygo, and
J. E. Lukens,  Quantum Superposition of Distinct Macroscopic
States, Nature {\bf 406} 43, (2000). 

\bibitem{beau1} I. Chiorescu, Y. Nakamura, C. J. P. M. Harmans, and
J. E. Mooij, Coherent Dynamics of a Superconducting Flux Qubit,
Science {\bf 299 } 186 (2003).


\bibitem{squid}
Study of Macroscopic Coherence and Decoherence in the SQUID by
Adiabatic Inversion, Paolo Silvestrini and \me,
Physics Letters {\bf A280} 17-22  (2001);www.arxiv.org,
cond-mat/0004472. Also
Adiabatic Inversion in the SQUID, Macroscopic Coherence and
Decoherence,   Paolo Silvestrini and \me,  {\it Macroscopic Quantum
Coherence and Quantum
Computing}, pg.271,  Eds. D. Averin, B. Ruggiero and P.
Silvestrini, Kluwer Academic/Plenum, New York (2001) www.arxiv.org,
cond-mat/0010129.

\bibitem{cnot} Design of   Adiabatic Logic for a Quantum CNOT Gate,
Valentina Corato, Paolo Silvestrini, \me, and Jacek Wosiek,
 www.arxiv.org, cond-mat/0205514, Physics
Letters {\bf A309} 206 (2003). Adiabatic Evolution of a Coupled-
Qubit Hamiltonian,
V.Corato, P. Silvestrini, \me, and J. Wosiek 
cond-mat/0310386; Physical Review {\bf B 68}, 224508 (2003).


\bibitem{sim}Valentina Corato, Paolo Silvestrini,\me, and Jacek
Wosiek, to be published. We  use a C program developed from the
work of J. Wosiek by  summer students at the Max-Planck-Institute,
A.T. Goerlich and P. Korcyl of the Jagellonian University.

\bibitem{grav} Decoherence Rate of Mass Superpositions, \me,
   Acta Physica Polonica {\bf B27}, 1915 (1996).
\end{references}
\end{document}